\documentclass[onecolumn]{autart}

\usepackage{graphicx}          

\usepackage{amsfonts}
\usepackage{amsmath}
\usepackage{amssymb}
\usepackage{graphicx}
\usepackage{float}
\usepackage{tcolorbox}
\usepackage{booktabs}
\usepackage[mathscr]{eucal}
\usepackage[american,cuteinductors,smartlabels]{circuitikz}
\usepackage{amsfonts}
\usepackage{float}
\usepackage[colorlinks,linkcolor=blue]{hyperref}
\usepackage[utf8]{inputenc}
\usepackage{amsmath}
\usepackage{graphicx}
\usepackage{amssymb}
\usepackage{verbatim}
\usepackage{epsfig}
\usepackage[labelfont=bf]{caption}
\usepackage{enumerate}
\usepackage{epstopdf}
\usepackage{subcaption}
\usepackage{siunitx}%
\usepackage[normalem]{ulem}

\newtheorem{lemma}{Lemma}
\newtheorem{proposition}{Proposition}
\newtheorem{corollary}{Corollary}
\newtheorem{remark}{Remark}
\newtheorem{assumption}{Assumption}

\def\begcen{\begin{center}}
	\def\endcen{\end{center}}

\newcommand{\col}{ \mbox{col} }

\def\calr{{\mathcal R}}
\def\calj{{\mathcal J}}

\def\call{{\mathcal L}}

\def\hal{{1 \over 2}}

\def\liminf{\lim_{t \to \infty}}

\def\L2{{\cal L}_2}
\def\L2e{{\cal L}_{2e}}

\def\rea{\mathbb{R}}

\def\diag{\mbox{diag}}

\def\begmat#1{\begin{bmatrix}#1\end{bmatrix}}
\def\begali#1{\begin{align}{#1}\end{align}}
\def\begalis#1{\begin{align*}{#1}\end{align*}}

\def\begequarr{\begin{eqnarray}}
	\def\endequarr{\end{eqnarray}}
\def\begequarrs{\begin{eqnarray*}}
	\def\endequarrs{\end{eqnarray*}}
\def\begarr{\begin{array}}
	\def\endarr{\end{array}}
\def\begequ{\begin{equation}}
	\def\endequ{\end{equation}}
\def\lab{\label}
\def\begdes{\begin{description}}
	\def\enddes{\end{description}}
\def\begenu{\begin{enumerate}}
	\def\begite{\begin{itemize}}
		\def\endite{\end{itemize}}
	\def\endenu{\end{enumerate}}

\def\lef[{\left[\begin{array}}
	\def\rig]{\end{array}\right]}

\def\begcen{\begin{center}}
	\def\endcen{\end{center}}
\def\begrem{\begin{remark}\rm}
	\def\endrem{\end{remark}}
\def\begassum{\begin{assumption}}
	\def\endassum{\end{assumption}}
\def\begassums{\begin{assumption*}}
	\def\endassums{\end{assumption*}}
\def\begassu{\begin{ass}}
	\def\endassu{\end{ass}}
\def\beglem{\begin{lemma}}
	\def\endlem{\end{lemma}}
\def\begcor{\begin{corollary}}
	\def\endcor{\end{corollary}}
\def\begfac{\begin{fact}}
	\def\endfac{\end{fact}}

\def\liminf{\lim_{t \to \infty}}

\def\L2e{{\cal L}_{2e}}

\def\rea{\mathbb{R}}

\def\diag{\mbox{diag}}

\def\col{\mbox{col}}
\def\hal{{1 \over 2}}

\def\diag{\mbox{diag}}

\def\begsubequ{\begin{subequations}}
	\def\endsubequ{\end{subequations}}

\usepackage{xcolor}

\graphicspath{{figs/}{}{img/}}

\begin{document}

\begin{frontmatter}

\title{Nonlinear Voltage Regulation of an Auxiliary Energy Storage of a Multiport Interconnection} 


\author[inh2]{Felipe Morales}\ead{felipe.morales@inh2.cl},   
\author[itam]{Rafael Cisneros}\ead{rcisneros@itam.mx},              
\author[itam]{Romeo Ortega}\ead{romeo.ortega@itam.mx},  
\author[utfsm]{Antonio Sanchez-Squella}\ead{antonio.sanchez@usm.cl}

\address[inh2]{Instituto Nacional del Hidr\'ogeno, INH2, 8940897 Santiago, Chile. }  
\address[utfsm]{Department of Electrical Engineering, Universidad T\'ecnica Federico Santa Mar\'ia, 8940897 Santiago, Chile. }
\address[itam]{Department of Electrical Engineering, ITAM,  01080 Mexico city, Mexico}


\begin{abstract}                          
In this article, we propose a nonlinear voltage control to ensure power exchange in a multiport interconnected system, which consists of a bidirectional DC-DC converter and generating-storing devices. The converter topology under consideration is two-stage, composed
of an interconnection of a buck with a boost converter. The motivation for this work is the explosive increase in the use of DC-DC converters due to the massification of renewable energies, electric vehicles powertrains, and energy storage systems, 
where fuel cells or batteries can be used as power backup or high-power support during transient phenomena. The converter's voltage  step-up and  step-down capabilities allow the use of supercapacitors with voltage limits that exceed those required by the load, thus enabling its use in a broader range of applications. The control design for this system does not correspond to that in standard applications involving power converters. As it is known, the latter consists of finding a control law such that the closed-loop system has an asymptotically stable equilibrium point fulfilling the voltage regulation objectives. Instead, in this application, the state does not tend to an equilibrium value in order for the system to be regulated. The converter voltage is regulated at desired some setpoint  whereas the other variables are only required to be bounded. To achieve a dynamic response that best adapts to changes in system demand and ensure stability over the defined wide operating range we propose a novel control strategy that exploits the partially cascaded structure of the system. Numerical and experimental results validate our approach.
\end{abstract}

\end{frontmatter}

\section{Introduction}
In various applications requiring energy storage, such as electric powertrains or backup energy systems, it is essential to have a DC-DC converter capable of managing the accumulator's energy and supplying it to the load according to its needs \cite{BESS,backup,backup_pv,EES_2,EES_PE,pbess}. A concrete example is found in the realm of electric vehicles with fuel cells, where an auxiliary energy source is crucial to support transient power phenomena required by the vehicle or battery storage systems. In this context, the DC-DC converter facilitates the change of state of the batteries, whether in charging or discharging, according to the system's demands \cite{fc_hybrid,fc_sc}. 

Within the broad spectrum of energy accumulators, supercapacitors have gained prominence due to their distinctive features, such as an extended lifespan, ability to handle high currents, longer operation cycles, and ease of maintenance. These devices find diverse applications, from smoothing energy to cover demand peaks without overloading the electrical grid, to dealing with short interruptions in the supply or providing momentary charging \cite{sc_app}. Moreover, they are employed to start engines in tanks, submarines, and locomotives. Recently, they have garnered significant interest due to their ability to rapidly absorb energy, making them suitable for regenerative braking applications in electric vehicles \cite{batsc,batsc_2}.

Given the versatility of supercapacitor applications, it is imperative to contemplate a circuit topology that allows managing the energy of this component across a wide range of voltages. For this reason, a DC-DC converter will be chosen with the capability to both reduce and increase the output voltage, as the supercapacitor's voltage varies with the state of charge, unlike other sources such as batteries. Additionally, this converter must enable bidirectionality of energy, allowing both the charging and discharging of the supercapacitor. Due to the need to react quickly to load variations, the control strategy must incorporate nonlinear characteristics that adapt optimally to the intrinsic dynamics of the converter and the supercapacitor. To attend this requirements, we consider a two-stage DC-DC converter composed of a cascade connection of a buck and a boost converter in a single design.  The converter is a two-port device that transmits energy from one of its port to the other. The control input permits to modulate the energy that flows  between its ports. It has to guarantee voltage regulation at one port regardless of the amount of power being transmitted an the direction of its flow. This feature allows using this topology in a wide variety of applications that require voltage regulation and working with large power values, such as electric vehicles or stationary energy storage systems.

For the control theoretic point of view the task we address in this paper is very challenging.  Indeed, as will be made precise below, our objective is to regulate only {\em part of the state} around some desired constant value, while we let the remaining states ``evolve in time" to meet the changing load demands. Consequently, it does not fit into the classical set-point regulation or trajectory tracking tasks. To solve this problem a novel approach is proposed in the paper. First, we introduce a decomposition of the system dynamics into two {\em partially cascaded} subsystems where each one of them contains the states that we want to regulate and includes independent control signals. Then, we select the latter to enforce the regulation objective for each subsystem via the introduction of a dynamic extension whose state plays the role of ``time-varying reference" for the remaining states. The design is completed proving that these two objectives are met ensuring boundedness of all signals.

The remainder of the paper is organized as follows. The description of the supercapacitor and DC-DC converter system analyzed in the paper is given in Section \ref{sec2}. In Section \ref{sec3}, we present two feedback representations of the overall system, one used to carried out the controller design, and the second illustrating an interesting robustness property of the proposed system. In Section \ref{sec4} we derive the controllers for each of the aforementioned subsystems, and in Section \ref{sec5} we present the overall control law and state the stability properties. Simulation and experimental results, which illustrate the closed-loop performance of the supercapacitor and DC-DC converter, are presented in Section \ref{sec6}. Finally, the paper is wrapped-up with concluding remarks and future work in  Section \ref{sec7}.

\noindent {\bf Notation.} $I_n$ is the $n \times n$ identity matrix. For $a \in \rea^n$, we denote $|a|^2:=a^\top a$. $\|x(t)\|_2$ is the $\mathcal{L}_2$-norm of the signal $x(t)$.

%
\section{System Description, Modelling and Control Objective}
\lab{sec2}
%

The topology of the system under study is depicted in Fig. 1. It consists of a two-stage dc-dc power converter  that processes energy, in a bidirectional way, from one of its ports to the other. At one end, there is a port connected to a supercapacitor---wich can be charged or discharged during the system operation. The port at the other end is connected to a current source $I_L$ which acts either as source ($I_L<0$) or load ($I_L>0$). When $I_L>0$, the supercapacitor operates in discharge mode as it provides its energy to the load.  Conversely, if $I_L<0$  the super capacitor operates in charge mode. 

\begin{figure}[h]
	\centering
	\includegraphics[scale=1]{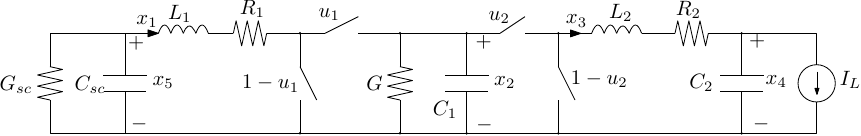}
	\caption{Considered topology.}
	\label{topology}
\end{figure}

\noindent The system is modeled by the following equations
\begin{subequations}\label{sys}
	\begin{align}
		L_1\dot{x}_1=&-R_1x_1 + x_5 -x_2u_1 \label{ec2}\\
		C_1\dot{x}_2=&{-Gx_2}+ x_1u_1 - x_3u_2 \label{ec1}\\
		L_2\dot{x}_3=&-R_2x_3-x_4+x_2u_2 \label{ec3}\\
		C_2\dot{x}_4=&x_3-I_L \label{ec4}\\
		C_{sc}\dot{x}_5=&-G_{sc}x_5-x_1.\label{ec5}
	\end{align}
\end{subequations}
where $x_1$ and $x_3$ correspond to the converter's inductor currents, $x_2$ and $x_4$ are the voltages across the conveter's capacitors. Resistors $R_1$ and $R_{2}$ are the inductors' parasitic resistances \footnote{We note that, if the supercapacitor model includes a losses series resistor, this can be simply added to $R_1$.}. Similarly, $G$ and $G_{sc}$ correspond to the parasitic parallel conductances of the capacitor $C_2$ and $C_{sc}$, respectively.\\

\begrem
\lab{rem1}
The system \eqref{sys} may be written in port-Hamiltonian form \cite{ESCVANORT} 
$$
Q\dot x=[\calj(u)-\calr]\nabla H(x)+gI_L,
$$
with the following definitions for the interconnection, damping and input matrices, 
$$
\calj(u) :=\begmat{0 & -u_1 & 0 & 0 & 0\\ u_1 & 0 & -u_2 & 0 & 0\\0 & u_2 & 0 & 0 & 0\\ 0 & 0 & 0 & 0 & 0\\0 & 0 & 0 & 0 & 0}=-\calj^\top(u),\;
\calr :=\begmat{R_1 & 0 & 0 & 0 & 0\\ 0 & G & 0 & 0 & 0\\0 & 0 & R_2 & 0 & 0\\ 0 & 0 & 0 & 0 & 0\\0 & 0 & 0 & 0 & G_{sc}} \geq 0,\;
g:=\begmat{0 \\ 0 \\ 0 \\ -1 \\ 0 },
$$
respectively, with $Q:=\diag\{L_1,C_1,L_2,C_2,C_{sc}\}$ and co-energy function $H(x)=\hal|x|^2$. As usual for port-Hamiltonian systems the power-balance equation
$$
\underbrace{\dot H}_{stored}=\underbrace{-x^\top \calr x}_{dissipated}  \underbrace{-x_4I_L}_{supplied\; power}
$$
is satisfied.
The power system might be composed of Fuel Cell, batteries, or supercapacitors, between other storage devices. In order to stretch the operation conditions and impose additional challenges to the controller we chose a supercapacitor. This device is in charge of storing the necessary energy to properly regulate the system voltages, while the load $I_L$ represents the supplied or absorbed power, resulting from a multi-port system, i.e., it supplies the current difference that the system must provide, accommodating specific demands. Unfortunately, the non-standard nature of the control task mentioned above, stymies the application of the well-established methods to design tracking or regulation controllers for port-Hamiltonian systems \cite{VANbook}. 
\endrem

\noindent Agreeing with its the physical operation, the following facts about the system are considered.

\begin{fact}\label{f1}\em
	Voltages $x_2,x_4$ and $x_5$ are positive.
\end{fact}

\begin{fact}\label{f2}\em 
	For $i\in\{1,2\}$, $u_i\in[u_m,u_M]$ where $u_m$ and $u_M$ are constants such that   $0<u_m<u_M\leq 1$.
\end{fact}

On the other hand, in view of the differences in time-scales between the motor and the other electrical components, we further assume the following.

\begin{assumption}\em
	\lab{ass1}
	The load current $I_L\in\mathbb{R}$ is {\em constant}.
\end{assumption}

\noindent {\bf Control Problem}
Design a model-based control strategy that ensures the following objectives.
\begin{enumerate}[{\bf O1}]
	\item Ensure {\em asymptotic convergence} of the capacitor voltage $x_4$ to the desired constant value $x^\star_{4}>0$. That is, ensure that 
	$$
	\liminf|x_4(t)-x_4^\star|=0.
	$$
	\item Guarantee {\em practical regulation} of the capacitor voltage $x_2$ around the desired value $x^\star_{2}>0$. That is, ensure that
	$$
	\liminf|x_2(t)-x_2^\star|\leq m,
	$$
	where the constant $m$ can be made {\em arbitrarily small} via the selection of the controller tuning gains.
	
	\item Ensure that these properties hold {\em for all} initial conditions and that all the signals remain {\em bounded}.\\	
\end{enumerate}

It is worth pointing out that, as indicated in the Introduction, the control task that we are confronting here is {\em not standard}. Indeed, it does not fit into the classical set-point regulation or trajectory tracking objectives but it aims at regulating only part of the state, {\em e.g.}, $(x_2,x_4)$ around some desired constant value, while we let the remaining states evolve in time to meet the load demands. In fact, there is no equilibrium point satisfying the control objectives $x_2=x_2^\star>0$ and $x_4=x_4^\star>0$. This can be confirmed from the equilibrium equation
\begin{align}\label{eq1}
0=[\calj( u^\star )-\calr]\nabla H(x^\star )+gI_L,
\end{align}
which yields $x_1^\star=-G_{sc}x_5^\star$ and $x_3^\star=I_L$. Furthermore, premultiplication of \eqref{eq1} by $x^{{\star}\top}$ produces
\begin{align*}
    0=&-(x^{\star})^\top \mathcal{R} x^{\star} -x_4^\star I_L\\
    =& -G (x_2^\star)^2-R_2I_L^2 -G_{sc}(1+R_1G_{sc}) (x_5^\star)^2-x_4^\star I_L.
\end{align*}
Clearly, for $I_L>0$, no real value of $x_5^\star$ satisfies the last equality.
%
\section{Two Suitable Decompositions of the Overall Systems}
\lab{sec3}
%
In this section we present two decompositions of the system \eqref{sys} as feedback interconnection of two subsystems. The first one is instrumental for the controller design that, as mentioned in the Introduction, is a novel approach to controller synthesis. On the other hand, the second decomposition reveals an interesting {\em robustness} property of the system, which is independent of the control signals.
\subsection{First decomposition for the controller design}
\lab{subsec31}
%

For the purpose of designing the control strategy, we find convenient to break the system down into two coupled subsystems as follows.
\begin{subequations}\label{sysdec}
	\begin{equation}	
		\lab{sig1}
		\Sigma_1:\left\{ \begin{aligned}
			L_2\dot x_3 = &-R_2x_3 - x_4 + x_2u_2\\
			C_2\dot x_4 = &  x_3 - I_L,
		\end{aligned}
		\right.
	\end{equation}
	\begin{equation}
		\lab{sig2}
		\Sigma_2\;:\;\left\{\begin{aligned}
			L_1\dot x_1 = & -R_1x_1 + x_5 -x_2u_1\\
			C_1\dot x_2 = & -Gx_2 +x_1u_1 - x_3u_2\\
			C_{sc}\dot x_5= & -G_{sc}x_5-x_1
		\end{aligned}\right.
	\end{equation}
\end{subequations}

In Fig. 2, viewing the controls $u_1$ and $u_2$ and the constant $I_L$ as external signals, we show a block diagram representation of the overall system as a feedback interconnection of two subsystems, $\Sigma_1:x_2 \to x_3$, with states $(x_3,x_4)$, and  $\Sigma_2:x_3 \to x_2$, with states $(x_1,x_2,x_5)$. 

\begin{figure}[h]
	\centering
	\includegraphics[scale=1.1]{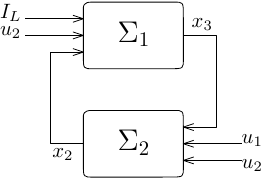}
	\caption{Feedback interconnection representation of the system \eqref{sys}}
\end{figure}

In the next section we exploit this partition to generate a signal $u_2$ such that---when replaced in the subsystem $\Sigma_1$---it will ensure {\em global regulation} of $x_4$. Then, replacing this control into subsystem $\Sigma_2$, we will define a control $u_1$ that guarantees {\em practical stability} of $x_2$. 
\subsection{Decomposition as feedback interconnection of two {\em passive} systems}
\lab{subsec32}
%
Before proceeding with the controller design using the decomposition of Fig. 2, we find convenient to uncover---in the lemma below---an interesting {\em robustness} property of the system \eqref{sys}, which is {\em independent} of the choice of the control signals $u_1$ and $u_2$.

\begin{lemma}
	\lab{lem1}\em
	The system \eqref{sys} admits the representation of Fig. 3 where, {\em for all} control signals $u_1$ and $u_2$, the operator $\bar \Sigma_1^M: v \to y$ is {\em output strictly passive} and the operator $\bar \Sigma_2: -y \to x_2$ is a strictly positive real (SPR) transfer function, hence it is {\em strictly passive}. As an immediate consequence of the Passivity Theorem \cite[Theorem 5.1]{DESVIDbook} we have that the closed-loop system operator is $\call_2$-stable. That is, there exists positive constants $\gamma$ and $\beta$ such that 
	$$
	\|y(t)\|_2 \leq \gamma \left\|{t \over u_2(t)}\right\|_2+ \beta.
	$$
\end{lemma}

\begin{figure}[h]
	\centering
	\includegraphics[scale=1.1]{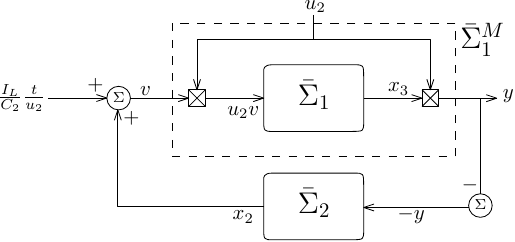}
	\caption{Representation of the system \eqref{sys} as a feedback interconnection of two {\em passive} systems.}
	\label{rep1}
\end{figure}

\begin{pf}
	Applying Laplace transform to the equations \eqref{sig1} of the subsystem $\Sigma_1$ we get
	\begalis{
		(L_2 s+R_2)X_3(s) &=-X_4(s) +W(s)\\
		C_2 s X_4(s)&=X_3(s)-{I_L \over s},
	}
	where we defined the signal $w:=x_2u_2$ and recalled that $I_L$ is a constant. Some simple manipulations of these equations show that
	$$
	X_3(s)={s \over L_2 s^2+R_2 s +{1 \over C_2}}\left[W(s)+{I_L \over C_2 s^2}\right].
	$$
	This relation is represented in the time domain in the forward path of Fig. \ref{rep2}. We make at this point three observations. First, that the transfer function ${s \over L_2 s^2+R_2 s +{1 \over C_2}}$ is SPR. Second, that the operator  $\bar \Sigma_1$  is an LTI system whose transfer function is precisely the aforementioned one. Finally, that it is possible to move the signal ${I_L \over C_2 s^2}$---whose inverse Laplace transform is ${I_L \over C_2}t$---outside of the feedback loop as done in Fig. \ref{rep1}. 
	
	Now, we recall \cite[Exercise 5]{DESVIDbook} where it is proven the pre- and posmultiplying a passive system by the same signal does no affect its passivity properties. Consequently, the operator $\bar \Sigma_1^M:v \to y$ of Fig. \ref{rep2} is passive. 
	
	\begin{figure}[h]
		\centering
		\includegraphics[scale=1.1]{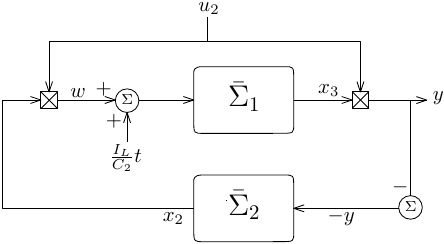}
		\caption{Alternative representation of the system \eqref{sys}}
		\label{rep2}
	\end{figure}
	
	\noindent To complete the proof we define the total co-energy of the system $\Sigma_2$ as
	$$
	H_2(x_1,x_2,x_5)=\hal \Big[L_1x_1^2+C_1x_2^2+C_{sc}x_5^2\Big],
	$$
	and compute its power-balance equation
	$$
	\dot H_2=-d_2-x_2x_3u_2=-d_2+x_2 \cdot(- y),
	$$
	where we recalled the definition of $y$ of Fig. 3 and defined the dissipation function
	$$
	d_2:=R_1x_1^2+G x_2^2+G_{sc}x_5^2 \geq 0.
	$$
	The output strict passivity claim of the operator $\bar \Sigma_2:(-y) \to x_2$ follows integrating the power balance equation.
\end{pf}

\begrem
\lab{rem2}
Notice that if $I_L=0$ the passivity of the operator $x_2 \to x_3u_2$ follows directly from the systems power-balance equation, differentiating the systems co-energy  
$$
H_1(x_1,x_2,x_5)=\hal \Big[L_2x_3^2+C_2x_4^2\Big],
$$
to get 
$$
\dot H_1=-R_2x_3^2+x_2\cdot(x_3u_2).
$$
\endrem

%
\section{Controller Design}
\lab{sec4}
%
In this section we carry out the design of a controller that achieves the control objective stated in Section \ref{sec2}. As indicated above we rely on the decomposition of the overall system \eqref{sys} described in Subsection \ref{subsec31} and proceed by deriving first the controller for the subsystem $\Sigma_1$ and then, complete the design proposing a controller for the subsystem $\Sigma_2$. Each one of these developments is described in a separate subsection below. 
\subsection{Controller for the subsystem $\Sigma_1$}
\lab{subsec41}
%
For this subsystem we design the control law for $u_2$ that ensures global regulation of $x_4$, with all signals bounded. This result is presented in the following lemma. 
\begin{lemma}\em
	\lab{lem2}
	Fix the desired constant reference for $x_4$ as $x_4^\star>0$. Consider $\Sigma_1$ given in \eqref{sig1} in closed loop with the dynamic state feedback
	\begsubequ
	\lab{con1}
	\label{s1}
	\begin{align}
		\lab{dotx3}
		\dot x_3^{\tt{ref}} &= -\frac{\kappa_{2}}{C_2} (x_3 - I_L) -\kappa_{3} \tilde x_4\\
		\lab{u2}	
		u_2 &=\frac{1}{x_2}\left[x_4+R_2 x_3^{\tt{ref}} - \kappa_{1} \tilde x_3 -L_2\kappa_{3}\tilde x_4 -\frac{L_2\kappa_{2}}{C_2}(x_3 - I_L) \right],
	\end{align}
	\endsubequ
	where we have defined the error signals
	\begalis{
		\tilde x_3&:=x_3-x_3^{\tt{ref}}\\
		\tilde x_4 &:=x_4-x_4^\star,
	}
	and for $i\in\{1,2,3\}$, $\kappa_{i}>0$. Then, for all states initial conditions,  $\lim_{t\to\infty} |\tilde x_4(t)|=0$ with $x_3(t) \in\mathcal{L}_\infty$ and  $x_3^{\tt{ref}}(t) \in\mathcal{L}_\infty$.
\end{lemma}
\begin{pf}
	Substituting $u_2$ into the first equation of $\Sigma_1$ yields
	\begalis{
		L_2\dot x_3 &= -R_2 \tilde x_3 - \kappa_{1} \tilde x_3 - L_2\kappa_{3}\tilde x_4-{L_2\kappa_2 \over C_2}( x_3-I_L)\\
		=& -(R_2+\kappa_{1})\tilde x_3+L_2 \dot  x_3^{\tt{ref}}
	}
	where we have used \eqref{dotx3} to get the second identity. The equation above may be written as
	$$
	L_2\dot{\tilde x}_3= -(R_2+\kappa_{1})\tilde x_3,
	$$
	from which we conclude that $\tilde x_3(t)\to 0$. To proceed with the proof we introduce a change of coordinates
	\begequ
	\lab{z}
	z:=-{1 \over k_3}(x_3^{\tt{ref}}+k_2 \tilde x_4).
	\endequ 
	Further, we notice that replacing the second equation of \eqref{sig1} in  \eqref{dotx3} we have that
	\begequ
	\lab{dotx3ref}
	\dot x_3^{\tt{ref}}=-k_2 \dot x_4 - k_3 \tilde x_4.
	\endequ
	Taking the derivative of $z$ in \eqref{z} we get
	\begalis{
		\dot z &= -{1 \over k_3}(\dot x_3^{\tt{ref}}+k_2 \dot{\tilde x}_4)=\tilde x_4,
	}
	where we used \eqref{dotx3ref} to get the second identity. Next, we carry out the following calculations
	\begin{align*}
		C_2\dot x_4=C_2\dot{\tilde x}_4=&x_3-I_L\\
		=& \tilde x_3 + x_3^{\tt{ref}} - I_L\\
		=&  \tilde x_3-\kappa_{2} \tilde x_4 - \kappa_3 z -I_L\\
		=& -\kappa_{2} \tilde x_4 - \kappa_3 \left(z + \frac{I_L}{\kappa_3} \right)+ \tilde x_3\\
		=&-\kappa_{2} \tilde x_4 -\kappa_3 \tilde z +\tilde x_3
	\end{align*}
	where we used \eqref{z} in the third identity and defined $\tilde z:= z+\frac{I_L}{\kappa_3}$ for the last equation. The proof of the claim is completed defining the state vector $\chi:=\col(\tilde x_3,\tilde x_4, \tilde z)$, whose dynamics is given by the asymptotically stable LTI system
	$$
	\dot \chi= \begin{bmatrix} -\frac{1}{L_2}(R_2 + \kappa_{1}) &0&0\\ \frac{1}{C_2}& -\frac{\kappa_2}{C_2} & -\frac{1}{C_2}\kappa_3 \\ 0 & 1 &0\end{bmatrix}\chi.
	$$
\end{pf}

\begrem
\lab{rem3}
As mentioned in the Introduction the dynamic extension $x_3^{\tt{ref}}$ serves as a ``time-varying" reference for the state $x_3$ that is selected in such a way as to ensure that the main regulation task $\tilde x_4(t) \to 0$ is effectively achieved. 
\endrem
\subsection{Controller for the subsystem $\Sigma_2$}
\lab{subsec42}
%
In this subsection we design the control law for $u_1$, which combined with the expression of $u_2$ given in Lemma \ref{lem2}, ensures that $x_2$ satisfies a practical stability property, with all signals bounded. This result is presented in the following lemma that, to simplify the proof, is presented in a {\em non-implementable} form---that is, involving open-loop time derivatives of signals. However, in a corollary after the proof, it is shown that this non-implementable form actually admits a standard dynamic state feedback form without open-loop time derivatives. 
\begin{lemma}
	\label{lem3}\em
	Fix the desired constant reference for $x_2$ as $x_2^\star>0$. Consider the subsystem $\Sigma_2$ given in \eqref{sig2} in closed loop with 
	\begsubequ
	\lab{conu1}
	\label{s2}
	\begin{align}
		\label{u1}
		u_1&=\frac{1}{x_2}\left[x_5+\kappa_4 \tilde x_1+w \right]\\
		\label{x1ref}	
		x_1^{\tt{ref}}& =- \kappa_5 \tilde x_2 \\
		\label{w}
		w &= -L_1\dot{x}_1^{\tt{ref}}-R_1x_1^{\tt{ref}},
	\end{align}
	\endsubequ
	where we defined the error signals
	\begalis{
		\tilde x_1&:=x_1-x_1^{\tt{ref}}\\
		\tilde x_2 &:=x_2-x_2^\star,
	}
	with $\kappa_{4}>0$ and $\kappa_5$ a gain {\em switching sign} such that $\kappa_5 x_1 \geq 0$.\footnote{See Subsection \ref{subsec43} for  further details on the choice of $\kappa_5$} Then, for all states initial conditions, the voltage error $\tilde x_2$ is {\em ultimately bounded}, that is,
	\begequ
	\lab{ultbou}
	\liminf|\tilde x_2(t)|\leq \frac{\Phi_M}{G+\kappa_5u_m},
	\endequ
	where $\Phi_M\in\mathbb{R}_+$ is a constant, and	the signals $x_1(t) \in\mathcal{L}_\infty$ and $x_5(t) \in\mathcal{L}_\infty$
\end{lemma}	
\begin{pf}
	Substituting $u_1$ into the first eq. of $\Sigma_2$ yields
	\begalis{
		L_1 \dot x_1&=-R_1x_1+x_5 -x_2u_1  \nonumber\\
		&=-R_1x_1- \kappa_4 \tilde x_1 - z \\
		&=-R_1x_1- \kappa_4 \tilde x_1  +L_1\dot{x}_1^{\tt{ref}}+R_1x_1^{\tt{ref}},
	}
	which may be written as
	$$
	L_1\dot{\tilde x}_1= -(R_1+\kappa_4)\tilde x_1.
	$$
	from which we conclude that $\tilde x_1(t)\to 0$. 	
	
	Consider now the error signal $\tilde x_2$. From the second equation in \eqref{sig2} we can do the following calculations 
	\begin{align*}
		C_1\dot x_2=C_1\dot{\tilde x}_2=&-Gx_2 +  x_1u_1 - x_3u_2\\
		=&-G  x_2 \pm Gx_2^\star  +x_1u_1 - x_3u_2 \\
		=&- G\tilde x_2  + (\tilde x_1+x_1^{\tt{ref}})u_1  - x_3u_2 -Gx_2^\star \\
		=& - (G+u_1\kappa_5) \tilde x_2  	+ \tilde x_1u_1 - x_3u_2-Gx_2^\star \\
		=& - (G +u_1\kappa_5) \tilde x_2  +\phi .
	\end{align*} 
	where we used \eqref{x1ref} to get the fourth  identity and defined in the last one the ``perturbation'' term 
	$$
	\phi:= \tilde x_1u_1 - x_3u_2-Gx_2^\star.
	$$  
	
	Let us define the signal
	\begin{align}
		\Phi:= |\tilde x_1| + |x_3| + |Gx_2^\star |,\label{Phi}
	\end{align}
	that, in view of Lemma \ref{lem2} and the analysis of $\tilde x_1$ above we conclude that it satisfies  $\Phi(t) \leq \Phi_M$, for some constant $\Phi_M$. From the definition of $\phi$ above we have that
	\begalis{
		|\phi| & = | \tilde x_1u_1 - x_3u_2-Gx_2^\star  | \\
		& \leq | \tilde x_1u_1| + |x_3u_2|+ |Gx_2^\star|\\
		& \leq \Phi \leq \Phi_M. 
	}
	where we used the fact that $u_i(t) <1,\;i=1,2,$ in the last inequality.
	
	We proceed now to prove that $\tilde x_2$ is bounded. To do so, we consider the quadratic function $W(\tilde x_2):= \frac{C_1}{2}\tilde x_2^2$, whose time derivative satisfies 
	\begin{align*}
		\dot V &= -G\tilde x_2^2  - \kappa_5 u_1 \tilde x_2^2 + \tilde x_2 \phi\\
		&\leq -G|\tilde x_2|^2 - \kappa_5 u_m |\tilde x_2|^2   + |\tilde x_2| \Phi \\
		&\leq  -[(G +\kappa_5 u_m) |\tilde x_2|-\Phi_M] |\tilde x_2| 			   
	\end{align*}
	which is negative whenever 
	$$
	|\tilde x_2|\geq \frac{\Phi_M}{G+\kappa_5 u_m}.
	$$ 
	The latter inequality provides the claimed ultimate bound \eqref{ultbou} of the regulation error $\tilde x_2$. 
	
    To finalize the proof it only remains to show that all signals are bounded. For, we see from \eqref{x1ref} that boundedness of $\tilde x_2$ implies that $x_1^{\tt{ref}}(t)\in\mathcal{L}_\infty$. Since we showed above that $\tilde x_1(t)\to 0$ this proves that $x_1(t)\in\mathcal{L}_\infty$, and from \eqref{ec5} we also conclude that $x_5$ is bounded.
\end{pf}

\begrem
\lab{rem4}
As indicated in the formulation of Lemma \ref{lem3}, for reasons explained in Subsection \ref{subsec43}, the gain $\kappa_5$ switches sign whenever $x_1$ does. To ensure that the bound in \eqref{ultbou} is well-defined it is necessary to impose the constraint 
\begequ
\lab{boug}
\kappa_5 \geq -{G\over u_m}+\epsilon,
\endequ
for some $\epsilon >0$. On the other hand, if $x_1$ does no change sign it is possible to reduce the size of the residual set increasing the gain $\kappa_5$.
\endrem
\subsection{An implementable realization of the controller of $\Sigma_2$}
\lab{subsec43}
%
In this subsection we present, as a corollary, an alternative form of the controller \eqref{conu1} that does not involve the noise-sensitive operation of open-loop time differentiation of signals. To present this result we need to take into account the following facts.
\begenu[{\bf F1}]
\item The current $x_1$ may take positive or negative values. Indeed, when the supercapacitor is charging $x_1<0$, while it is positive if it is discharging.
\item Due to physical constraints it is possible to define constants  $x^{\min}_1<0$ and $x^{\max}_1>0$ such  that $x^{\max}_1\geq x_1(t) \geq x^{\min}_1$ for all $t \geq 0$.
\endenu
\begin{corollary}\em  
	\lab{cor1}
	Assume the tuning gain $\kappa_5(t)$ is selected such that \eqref{boug} and
	\begali{
		\nonumber
		\kappa_5(t) & \geq \frac{C_1 }{L_1} {x_2(t) \over x^{\max}_1}+ \epsilon\quad if\;x_1(t) \geq 0\\
		\lab{assk5}
		\kappa_5(t) & \leq \frac{C_1 }{L_1} {x_2(t) \over x^{\min}_1}- \epsilon \quad if\;x_1(t) <0
	}
	are satisfied for some $\epsilon > 0$. Under this condition, the controller \eqref{conu1} admits the standard dynamic state-feedback realization
	\begsubequ
	\lab{con2}
	\begali{
		\lab{dd}
		\dot x_1^{\tt{ref}} &= -\frac{\kappa_5}{C_1} (x_1u_1 -x_3u_2-Gx_2)\\
		\lab{du}
		u_1 &= \left( x_2 - \frac{L_1 \kappa_5}{C_1}x_1  \right)^{-1}\left[x_5- R_1x_1^{\tt{ref}} - \kappa_4  \tilde x_1 - \frac{L_1}{C_1}  \kappa_5 (x_3u_2+Gx_2 )  \right],
	}
	\endsubequ
	with $u_2$ given in \eqref{u2}.
\end{corollary}
\begin{pf}
	From \eqref{x1ref} we get that
	\begin{align*} 
		\dot x_1^{\tt{ref}} =& -\kappa_5 \dot x_2 \\
		=& -\frac{\kappa_5}{C_1} (x_1u_1 -x_3u_2-Gx_2),
	\end{align*}
	where we used \eqref{ec1} in the second identity. Replacing the latter in \eqref{conu1} yields the alternative expression of $u_1$
	\begin{align}\label{cont_1}
		u_1&=\frac{1}{x_2}\left[x_5 - \kappa_4 \tilde x_1 + \frac{L_1 \kappa_5}{C_1}  (x_1u_1 -x_3u_2-Gx_2)  -R_1x_1^{\tt{ref}} \right], 
	\end{align}
	which, after multiplying both sides by $x_2$, and grouping terms may be written as
	\begin{align*}
  \left( x_2 - \frac{L_1 \kappa_5}{C_1}x_1  \right)	u_1 = x_5 - \kappa_4 \tilde x_1 - \frac{L_1\kappa_5}{C_1}\left(   x_3u_2+Gx_2\right) - R_1x_1^{\tt{ref}}
	\end{align*}
	The proof is completed invoking the condition \eqref{assk5} that ensures the left-hand term in parenthesis above is bounded away from zero.
\end{pf}



\begrem
\lab{rem5}
From \eqref{assk5} it is clear that $\kappa_5(t)$ must switch sign whenever the direction of the energy flow to the supercapacitor changes direction, ensuring $\kappa_5 x_1 \geq 0$.
\endrem
%
\section{Control Law for the Overall System}
\lab{sec5}
%
\noindent Based on Lemmata \ref{lem2} and \ref{lem3} and Corollary \ref{cor1}, we can state our main result, namely, the control law for the overall system \eqref{sys} that achieves the control objectives {\bf O1-O3} of Section \ref{sec2}.
\begin{proposition}\em
	\lab{pro1} 
	Fix the desired constant references for $(x_2,x_4)$ as $(x_2^\star,x_4^\star)$. Select the gain $\kappa_5$ such that the conditions \eqref{boug} and \eqref{assk5} are satisfied, where  $x_i \in [x^{\min}_i,x^{\max}_i],\;i=1,2$. 
	
	Consider the system \eqref{sys} in closed loop with \eqref{con1} and \eqref{con2}, where $\kappa_{i}>0,\;i\in\{1,\dots,4\}$. The following properties hold true.
	\begenu[{\bf P1}]
	\item $x_4$ is asymptotically regulated to its desired value, that is, 
	$$
	\lim_{t\to\infty} |\tilde x_4(t)|=0.
	$$
	\item $x_2$ is ultimately bounded around its desired value, that is,
	$$
	\liminf|\tilde x_2(t)|\leq \frac{\Phi_M}{G+\kappa_5u_m}.
	$$
	\item These properties are satisfied for all initial conditions.\\
	\item All signals remain bounded. 
	\endenu
\end{proposition}
%
\section{Simulation and Experimental Results}
\lab{sec6}
%

\subsection{Simulation Results} 

To evaluate the performance of the controllers designed for the system depicted in Fig. \ref{topology}, simulation software, namely Simulink and PLECS, was employed. In Simulink, the controllers specified by expressions \eqref{s1}, \eqref{s2} and \eqref{con2} were programmed  
while in PLECS, the topology of the studied system was modeled, with switches toggling at a frequency of \SI{10}{\kilo\hertz}. The values of the system components are shown in Table \ref{tab:parameter}, and the baseline values used for the simulations are presented in Table \ref{tab:controller}. The inputs signals $u_1$ and $u_2$ are modulated to obtain the corresponding switching signals which takes values in the set $\{0,1\}$.

\begin{table}[!ht]
	\centering
	\begin{tabular}{cc}
		\hline
		Parameter & Value\\ \hline \hline
		$L_1$ & \SI{10}{\milli\henry} \\
		$L_2$ & \SI{10}{\milli\henry} \\
		$R_1$ & \SI{113.2}{\milli\ohm}\\ 
		$R_2$ & \SI{100}{\milli\ohm}\\
		$C_1$ & \SI{8.8}{\milli\farad}\\
		$C_2$ & \SI{2.2}{\milli\farad}\\
		$C_{sc}$ & \SI{62.5}{\farad} \\
		$G$ & \SI{50}{\micro\siemens}\\
		$G_{sc}$ & \SI{200}{\micro\siemens} \\ \hline
	\end{tabular}
	\caption{System parameters used in the simulation.}
	\label{tab:parameter}
\end{table}

\begin{table}[!ht]
	\centering
	\begin{tabular}{cc}
		\hline
		Parameter & Value\\ \hline \hline
		$\kappa_{1}$ & 10 \\
		$\kappa_{2}$ & 1 \\
		$\kappa_{3}$ & 500\\
		$\kappa_{4}$ & 1\\
		$\kappa_{5}$ & 1.8 \\ \hline
	\end{tabular}
	\caption{Baseline values for the controllers used in the simulation.}
	\label{tab:controller}
\end{table}

\begin{figure}[!h]
	\centering
	\begin{subfigure}{0.40\textwidth}
		\centering
		\includegraphics[width=0.9\textwidth]{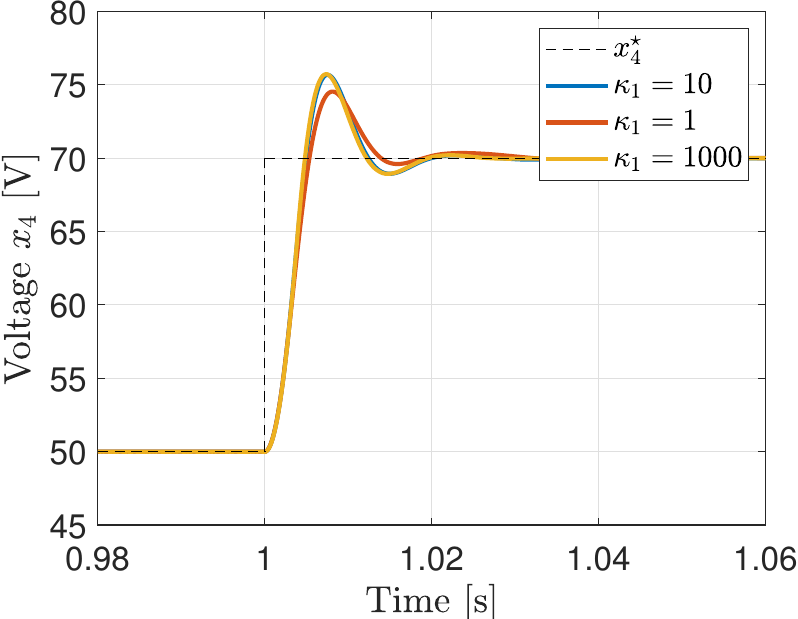}
		\caption{Different values of $\kappa_1$.}
		\label{k1}
	\end{subfigure}\hfil
	\begin{subfigure}{0.40\textwidth}
		\centering
		\includegraphics[width=0.9\textwidth]{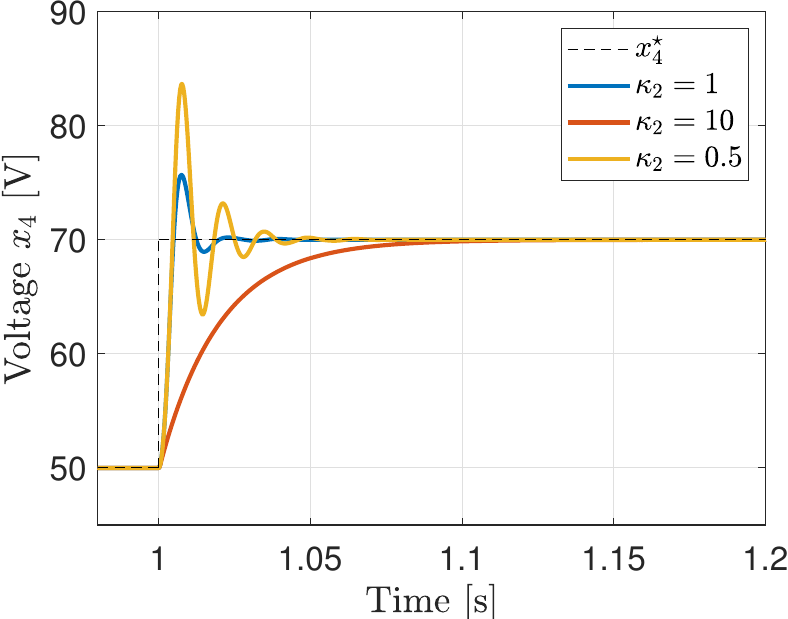}
		\caption{Different values of $\kappa_2$.}
		\label{k2}
	\end{subfigure}\\
	\begin{subfigure}{0.40\textwidth}
		\centering
		\includegraphics[width=0.9\textwidth]{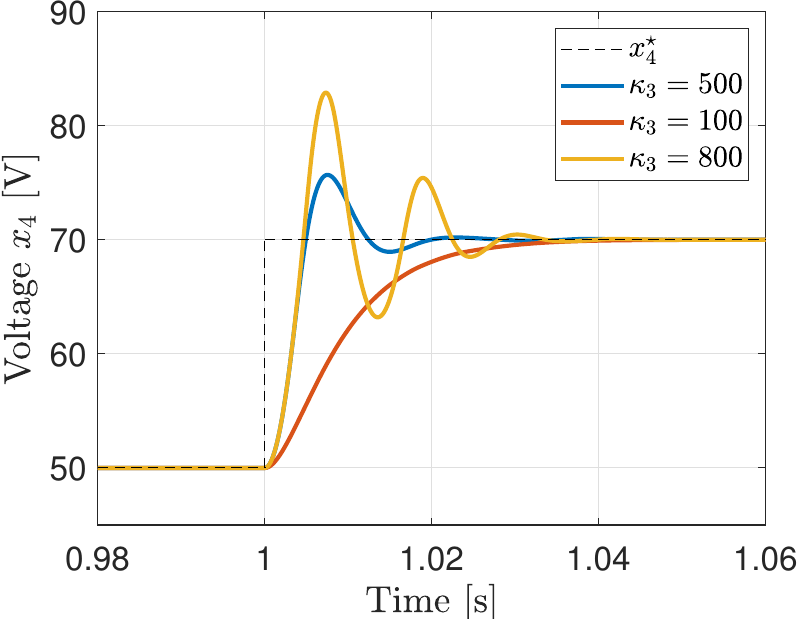}
		\caption{Different values of $\kappa_3$.}
		\label{k3}
	\end{subfigure}     
	\caption{System $\Sigma_1$: impact of the controller gains on $x_4$ when $x_4^\star$ varies in \textit{simulation}.}
	\label{x4}
\end{figure}

To test the controlled system and evaluate the dynamic effect of the tuning parameters, several simulations were conducted with different values of $\kappa_i$. Analyzing each subsystem separately, as presented in Fig. \ref{x4}. For subsystem $\Sigma_1$, a reference change was applied to $x_4^{\star}$ from 50 to \SI{70}{\volt}, adjusting the values of $\kappa_1$, $\kappa_2$, and $\kappa_3$, while keeping $\kappa_4$ and $\kappa_5$ at their baseline values from Table \ref{tab:controller}. It is noteworthy that during this simulation, the value of $I_L$ was held constant at \SI{5}{\ampere}. 

As depicted in Fig. \ref{x4}, the increment of $\kappa_1$ has the effect of enhancing the overshoot in the dynamic response and reducing the stabilization time. On the other hand, varying $\kappa_2$ alters the system's response type, transitioning from underdamped to overdamped, along with modifying the stabilization time. Similarly, $\kappa_3$ also influences these aspects. It is noteworthy that in all evaluated scenarios, the desired value for $x_4^{\star}$ is successfully reached and the stabilization times range between 40ms and 100ms.

\begin{figure}[!h]
	\centering
	
	\begin{subfigure}{0.40\textwidth}
		\centering
		\includegraphics[width=0.9\textwidth]{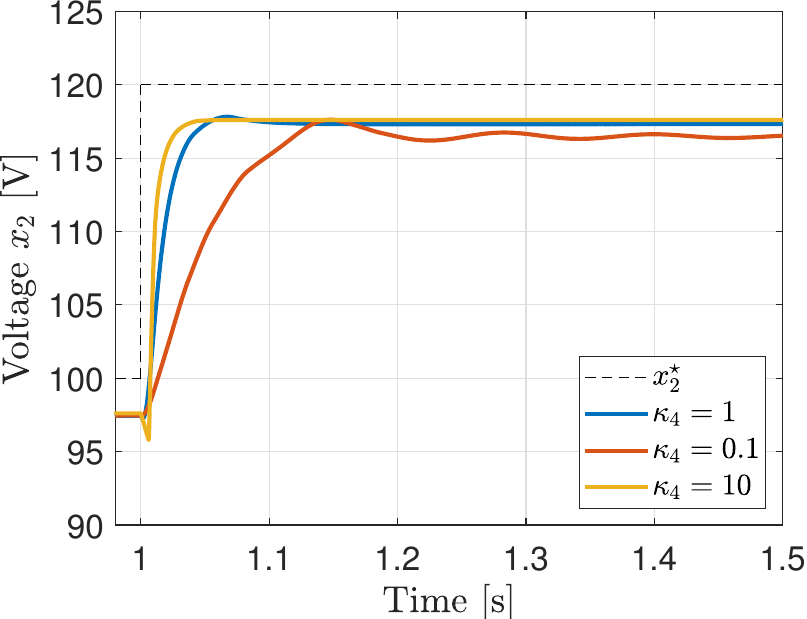}
		\caption{Different values of $\kappa_4$.}
		\label{k4}
	\end{subfigure}\hspace{10pt}
	\begin{subfigure}{0.40\textwidth}
		\centering
		\includegraphics[width=0.9\textwidth]{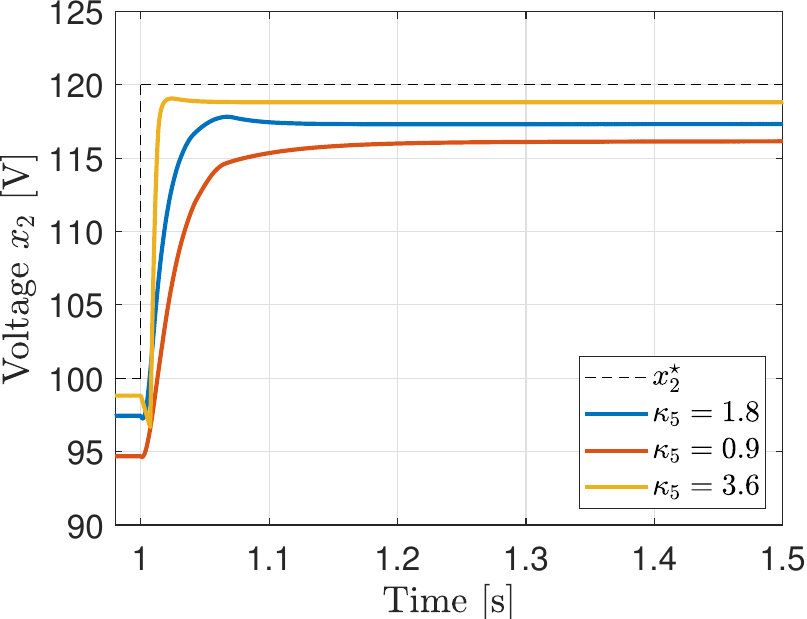}
		\caption{Different values of $\kappa_5$.}
		\label{k5}
	\end{subfigure}   
	\caption{System $\Sigma_2$: impact of the controller gains on  $x_2$ when $x_2^\star$ varies in \textit{simulation}.}
	\label{x2}
\end{figure}

In the case of subsystem $\Sigma_2$, the values of $\kappa_1$, $\kappa_2$, and $\kappa_3$ are kept at their baseline values, while the rest are modified to analyze their effect. During this simulation, the value of $x_2^{\star}$ changes from 100 to \SI{120}{\volt}, with $I_L$ maintained at a constant value of \SI{5}{\ampere} to assure the energy transfer, which is the main objective. Figure \ref{x2} illustrates that increasing $\kappa_4$ or $\kappa_5$ has the effect of decreasing the steady-state error from \SI{116}{\volt} to \SI{117.5}{\volt} in figure \ref{k4} and from \SI{116}{\volt} to \SI{119}{\volt} in figure \ref{k5}, for variable $x_2$ and also allows for a shorter signal stabilization time. The existence of steady-state error does not jeopardize the energy transfer and is consistent with the limits shown previously, where an increase in $\kappa_5$ leads to a decrease in steady-state error.

\begin{figure}[!h]
	\centering
	\includegraphics[width=0.45\textwidth]{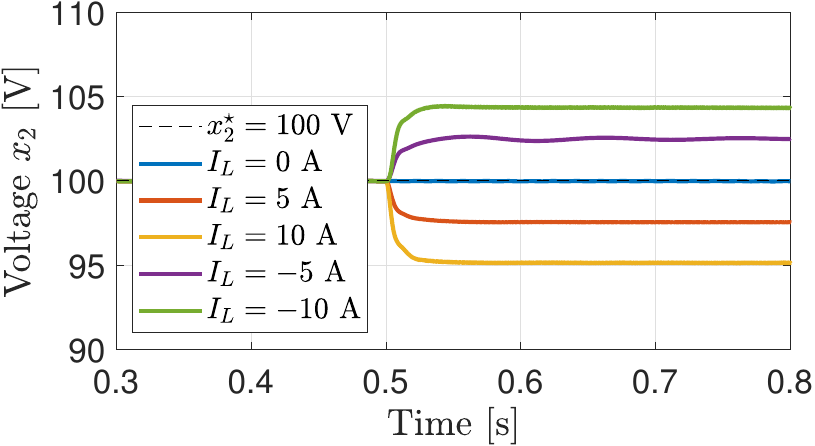}
	\caption{Regulation of $x_2$ for different values of $I_L$.}
	\label{il}
\end{figure}

Note that the load $I_L$ also affects system response since currents $x_1$ and $x_3$ depend on this value, altering the magnitude given by expression \eqref{Phi} and changing the steady-state bound of signal $x_2$. This is depicted in Fig. \ref{il}, where an increase in the absolute value of the current leads to an increase in steady-state error. Additionally, the sign of the current, i.e., whether the supercapacitor is in charging or discharging mode, determines whether the error $\tilde x_2$ is positive or negative. The latter should be taken into consideration during design and implementation to avoid exceeding the nominal voltage value of capacitor $C_1$. 

\begin{figure}
    \centering
    \includegraphics[width=0.5\textwidth]{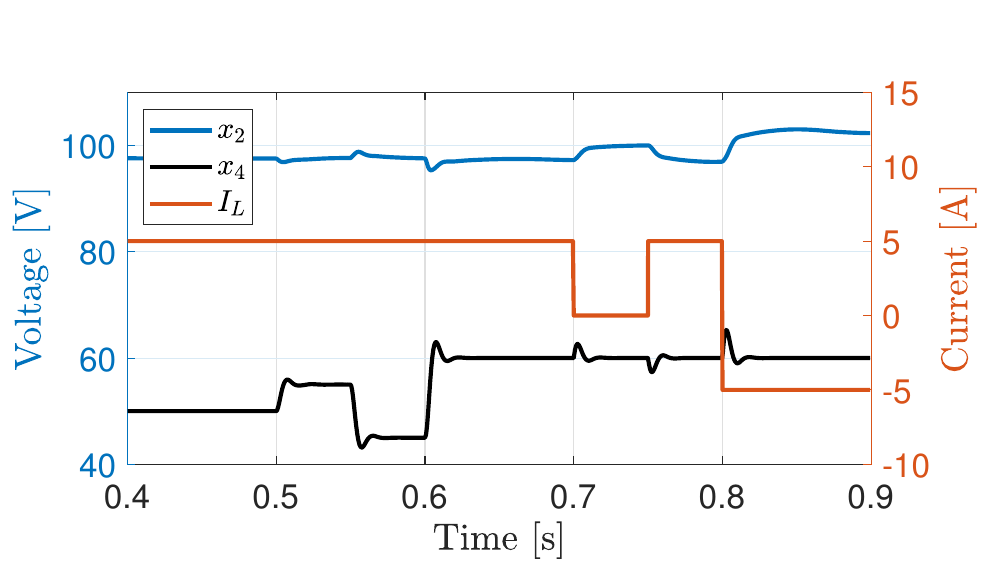}
    \caption{Voltage regulation when either $x_2^\star$ or $I_L$ vary.}
    \label{fig:full}
\end{figure}

\begin{figure}
    \centering
    \includegraphics[width=0.5\textwidth]{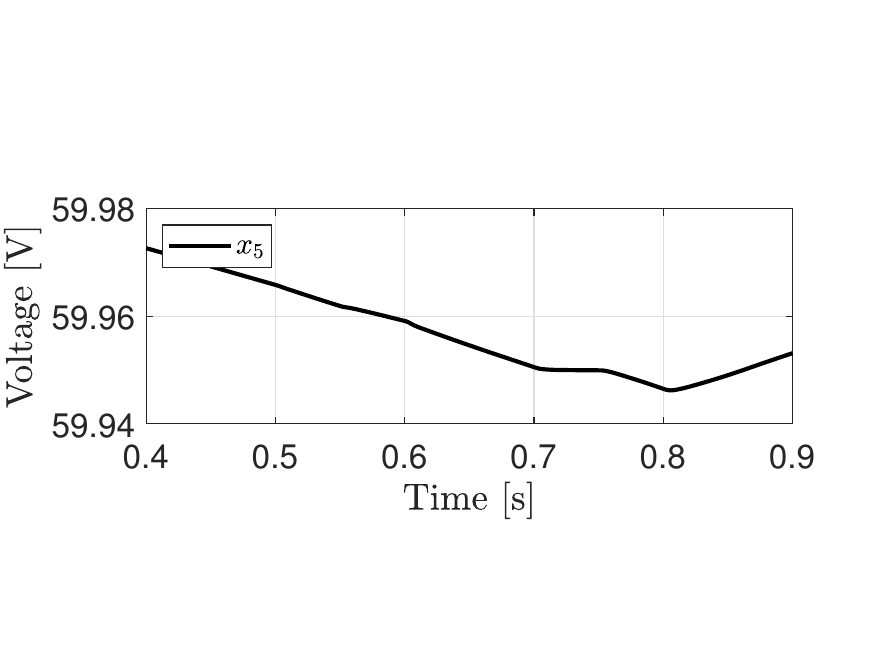}
    \caption{Voltage variation of $x_5$ when either $x_2^\star$ or $I_L$ vary}
    \label{fig:x5}
\end{figure}

In Fig. \ref{fig:full}, another simulation results are shown consisting in the following scenario. During the time interval $t=[0,\;0.6]$, voltage reference of $x_2$ and $I_L$ are both  held constant at $x_2^\star=100\mathrm{V}$ and  $I_L=5\mathrm{A}$, respectively. Meanwhile, reference $x_4^\star$ is stepwise varying, and taking values in the set $\{45\mathrm{V},\;50\mathrm{V},\;55\mathrm{V},\;60\mathrm{V} \}$. As it can be seen from the figure, $x_4$ tends to its reference whereas $x_2$ is bounded during this period of time. From $t=0.6$s to the end time of  simulation, $x_4^\star=60\;\mathrm{V}$ and $x_2^\star$ is newly kept constant at $\SI{100}\volt$  while $I_L$ varies in a stepwise manner, taking values in the set $\{\SI{-5}{\ampere}, \SI{0}{\ampere}, \SI{5}{\ampere}\}$. As before, output voltage  $x_4$ is regulated at any time. Some transients are present as a consequence of the current variation. We conclude that the designed controller effectively regulates, regardless of the load it is powering or its polarity. Also, in Fig. \ref{fig:x5}, the variation of the voltage of $x_5$ during the scenario described above is shown, where the periods of discharge or charging of the supercapacitor are observed, according to the current $I_L$.


\subsection{Experimental Results}

\begin{figure}[!t]
	\centering
	
	\begin{subfigure}{0.45\textwidth}
		\centering
		{\setlength{\fboxsep}{-2pt}%
			\setlength{\fboxrule}{2pt}%
			\fbox{\includegraphics[width=\textwidth]{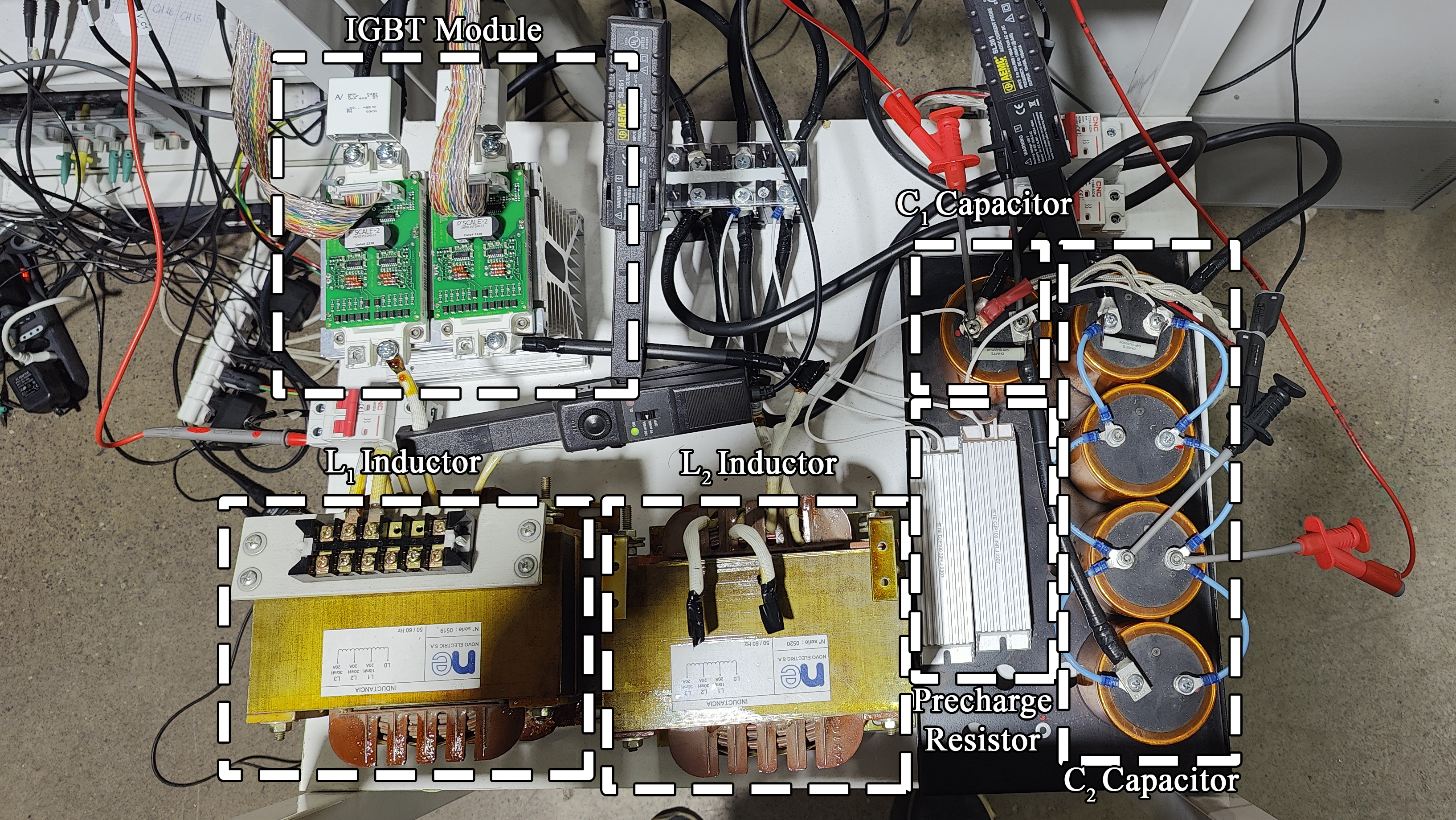}}}
		\caption{DC-DC Converter.}
		\label{fig:dcdc_montaje}
	\end{subfigure}     
	\hspace{2pt}
	\begin{subfigure}{0.45\textwidth}
		\centering
		{\setlength{\fboxsep}{-2pt}%
			\setlength{\fboxrule}{2pt}%
			\fbox{\includegraphics[width=\textwidth]{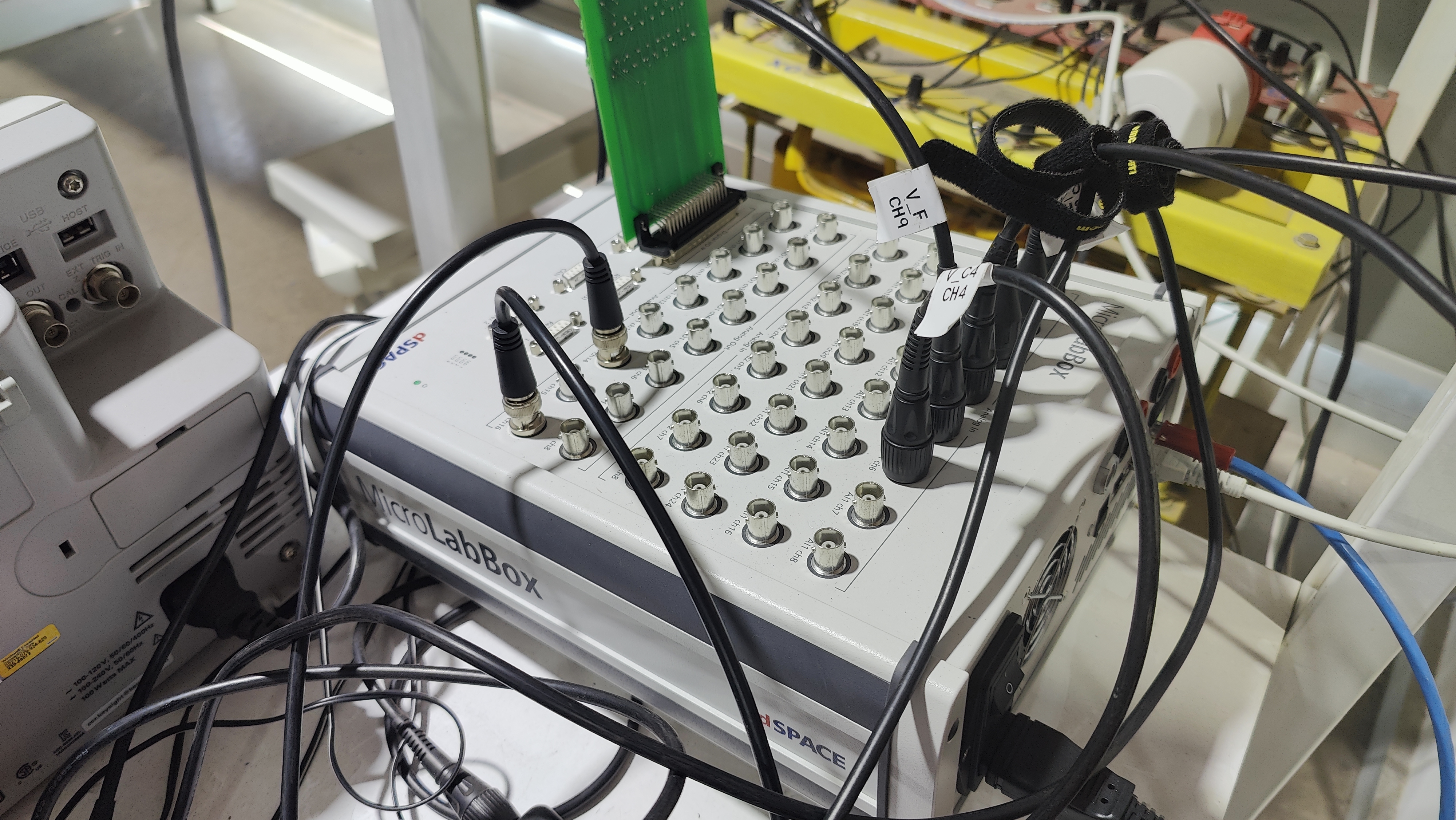}}}
		\caption{MicroLabBox Platform.}
		\label{fig:dspace_montaje}
	\end{subfigure}
	\hspace{200pt}
	\begin{subfigure}{0.45\textwidth}
		\centering
		{\setlength{\fboxsep}{-2pt}%
			\setlength{\fboxrule}{2pt}%
			\fbox{\includegraphics[width=\textwidth]{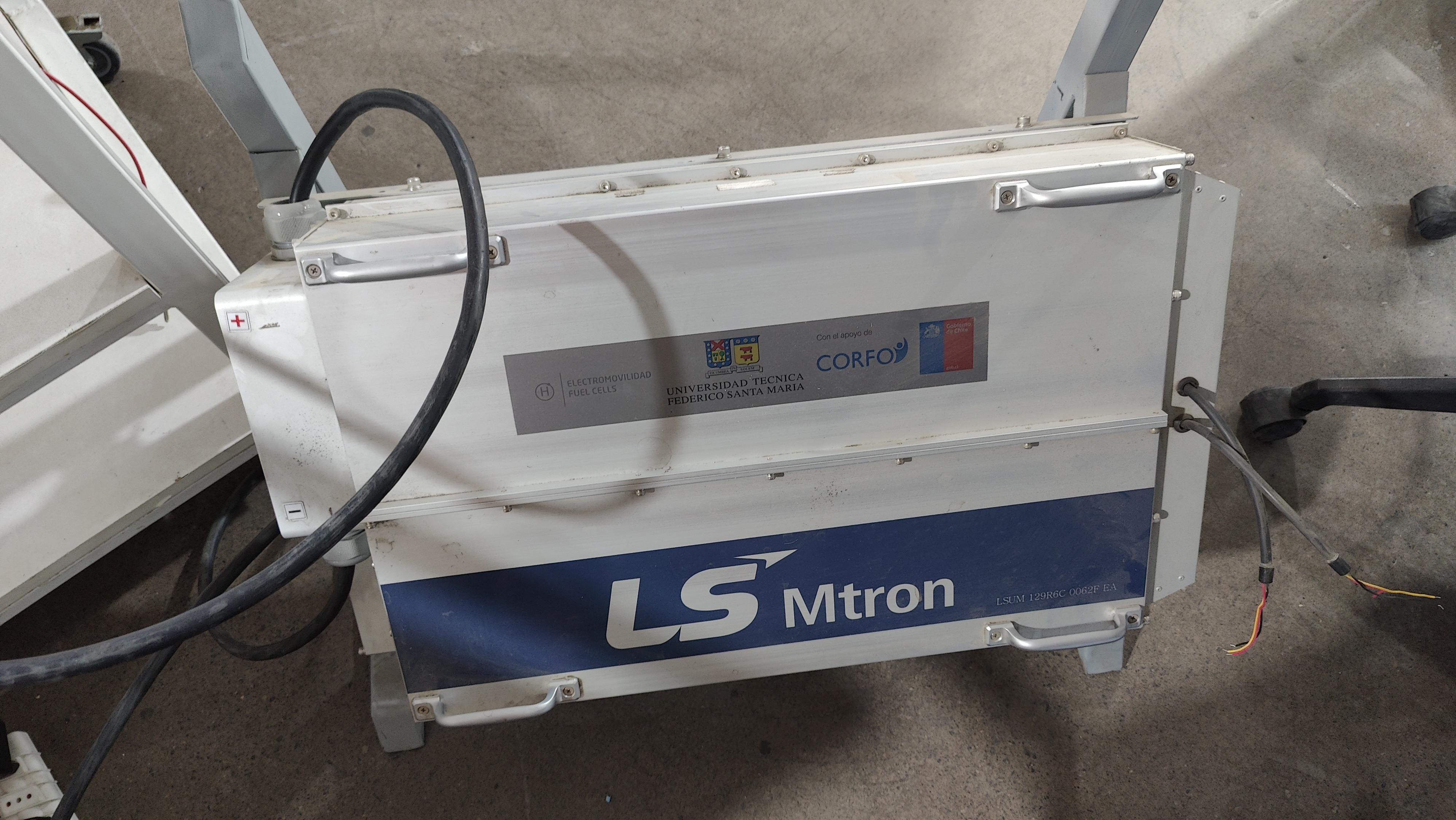}}}
		\caption{LSUM129 supercapacitor.}
		\label{fig:SC}
	\end{subfigure}   
	\hspace{2pt}
	\begin{subfigure}{0.45\textwidth}
		\centering
		{\setlength{\fboxsep}{-2pt}%
			\setlength{\fboxrule}{2pt}%
			\fbox{\includegraphics[width=\textwidth]{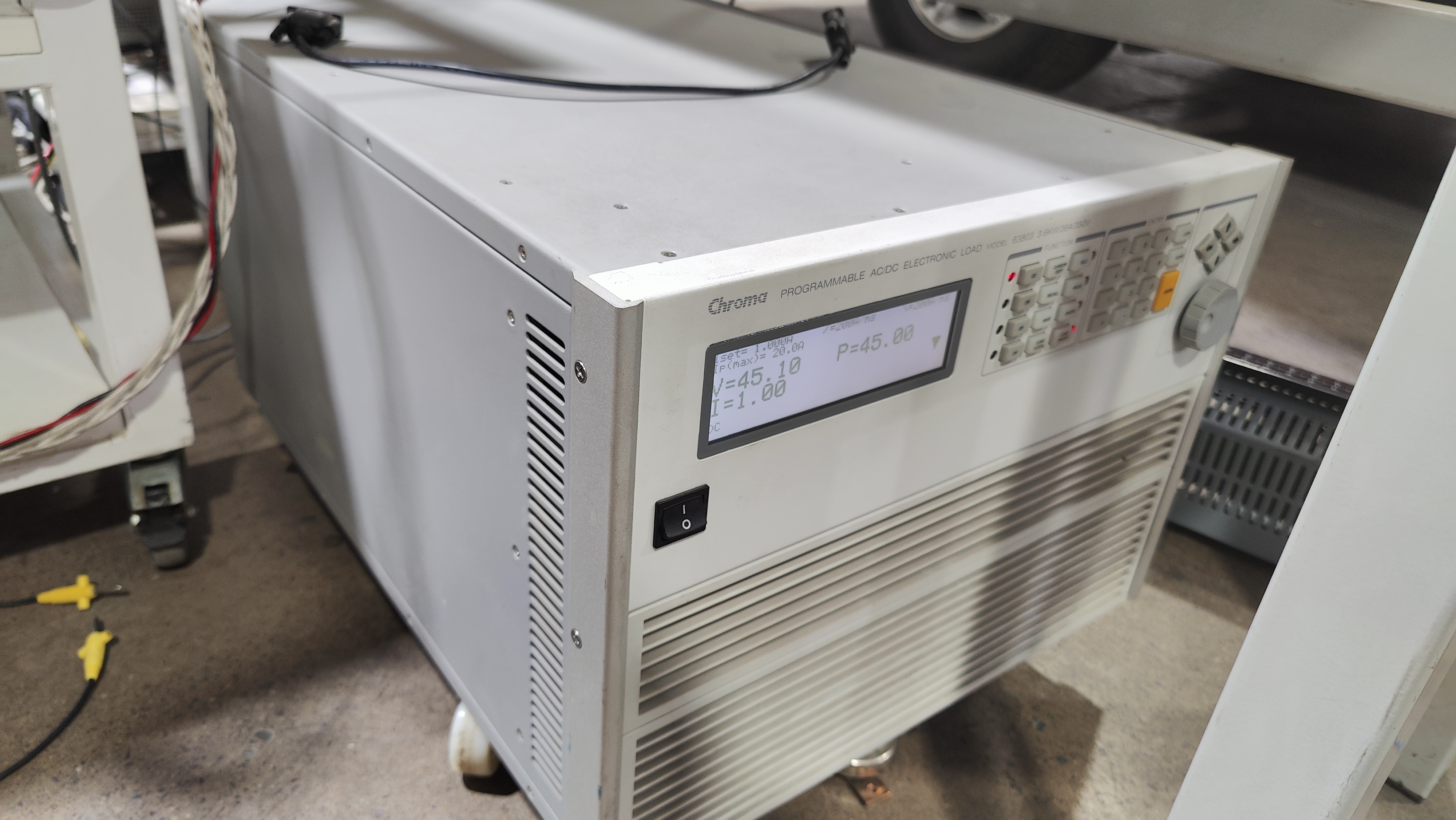}}}
		\caption{Chroma programmable DC Load.}
		\label{fig:load}
	\end{subfigure}  
	\caption{Testing platform used for experiments.}
	\label{testing}   
\end{figure}

To test the developed control strategy, a test platform was assembled, consisting of a DC-DC converter, a supercapacitor, and a programmable electric load. The parameter values for each component are listed in Table \ref{tab:medidos}. The controllers were programmed using the MicroLabBox platform from dSPACE, where the values of $\kappa_i$ used in the controllers are displayed in Table \ref{tab:controller_2}. In Fig. \ref{testing}, photographs of the test bench and the equipment used in the experimentation are displayed.

\begin{table}[!h]
	\centering
	\begin{tabular}{cccccc}\toprule
		Parameter & Value \\ \midrule \midrule
		$L_1$ & \SI{8.78}{\milli\henry} \\
		$L_2$ & \SI{8.55}{\milli\henry}  \\
		$C_1$ & \SI{2.19}{\milli\farad}  \\ 
		$C_2$ & \SI{7.6}{\milli\farad}  \\ 
		$R_1$ &   \SI{1.58}{\ohm} \\ 
		$R_2$ &   \SI{1.69}{\ohm} \\ 
		$G_{sc}$ & \SI{200}{\micro\siemens} \\
		$G$ & \SI{50}{\micro\siemens} \\
		$f_{sw}$ & \SI{10}{\kilo\hertz}  \\ \bottomrule
	\end{tabular}
	\caption{Parameter values used in the test platform.}
	\label{tab:medidos}
\end{table}

\begin{table}[!h]
	\centering
	\begin{tabular}{cc}
		\hline
		Parameter & Value\\ \hline \hline
		$\kappa_{1}$ & 1 \\
		$\kappa_{2}$ & 0.5 \\
		$\kappa_{3}$ & 10\\
		$\kappa_{4}$ & 1\\
		$\kappa_{5}$ & 1.8 \\ \hline
	\end{tabular}
	\caption{Baseline values for the controllers used in the experimentation.}
	\label{tab:controller_2}
\end{table}

During the experiments, the goal was to observe the effect of the $\kappa_i$ values when making a reference change in the controlled variables, along with the impact of the system's power supply load and the effect of $\kappa_5$ on the error in controlling variable $x_2$.

\begin{figure}[!h]
	\centering
	
	\begin{subfigure}{0.40\textwidth}
		\centering
		\includegraphics[width=0.9\textwidth]{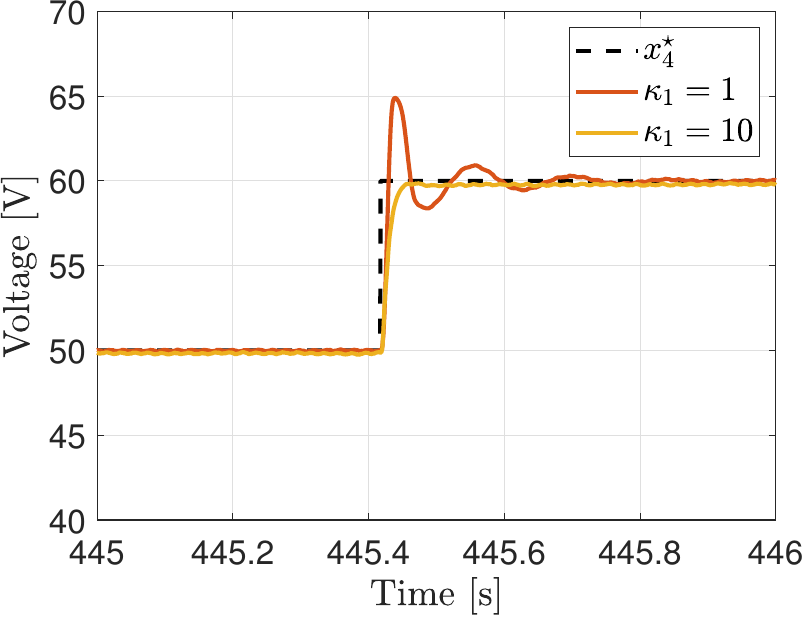}
		\caption{Different values of $\kappa_1$.}
		\label{x4_k1}
	\end{subfigure}\hspace{10pt}
	\begin{subfigure}{0.40\textwidth}
		\centering
		\includegraphics[width=0.9\textwidth]{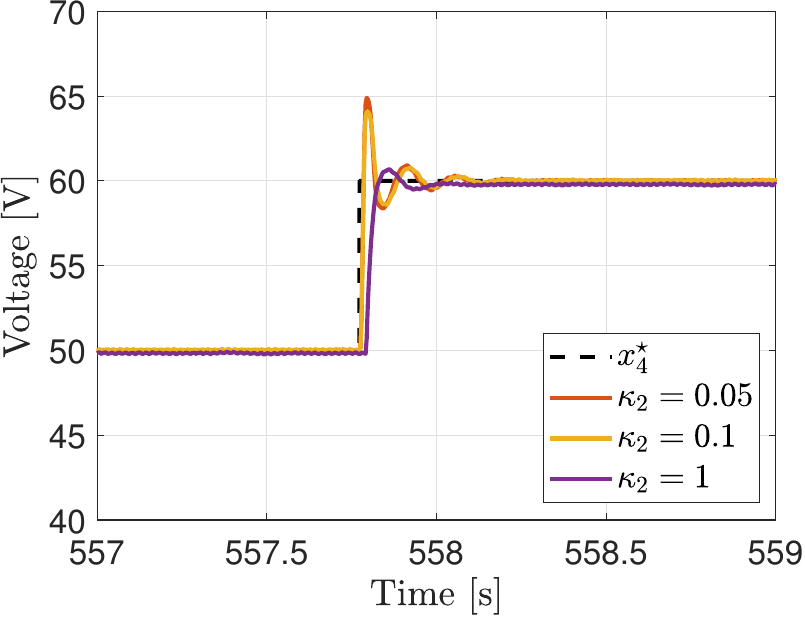}
		\caption{Different values of $\kappa_2$.}
		\label{x4_k2}
	\end{subfigure}   \\
    \begin{subfigure}{0.40\textwidth}
		\centering
		\includegraphics[width=0.9\textwidth]{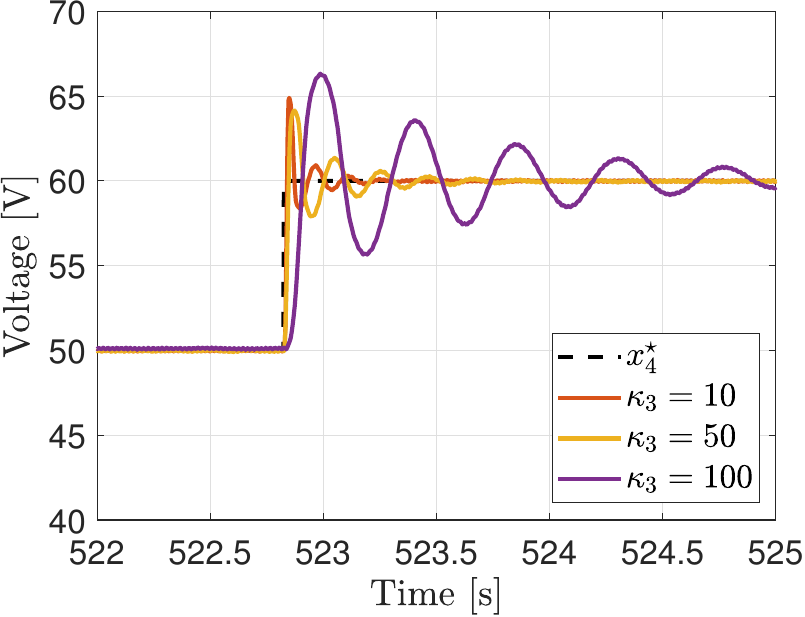}
		\caption{Different values of $\kappa_3$.}
		\label{x4_k3}
	\end{subfigure} 
	\caption{System $\Sigma_1$: impact of the controller gains on  $x_4$ when $x_4^\star$ varies in \textit{experimentation}.}
	\label{x_2}
\end{figure}

As despicted in Fig. \ref{x4_k1} when a 10V step was applied to $x_4^{\star}$, a notorious reduction of the damping is experienced when $\kappa_1=1$ with respect to the response when $\kappa_1=10$---just as expected from the simulation results. Increasing $\kappa_1$ impacts the steady state, which increases in 1.67\%.

%

For $\kappa_2$, three values were used, and it was observed that an increase in the value of $\kappa_2$ damped the controlled variable and allowed it to reach the desired value more quickly, stabilization time is reduced from 0.48s to 0.3s, this is shown in Fig. \ref{x4_k2}.

%
%
Figure \ref{x4_k3} illustrates the effect of $\kappa_3$ on the response to a change in the reference of $x_4^{\star}$. It is observed that a higher value of $\kappa_3$ directly influenced the oscillatory behavior of the signal, with lower damping and higher stabilization times. Based on the above results, it can be concluded that the subsystem $\Sigma_1$ is effectively controlled, achieving the desired objective. Although $\kappa_3$ is increase in 10 times, the system stabilizes in 5s, proving the robustness of the controller.

%
%
%

\begin{figure}[!h]
	\centering
	\includegraphics[width=0.45\textwidth]{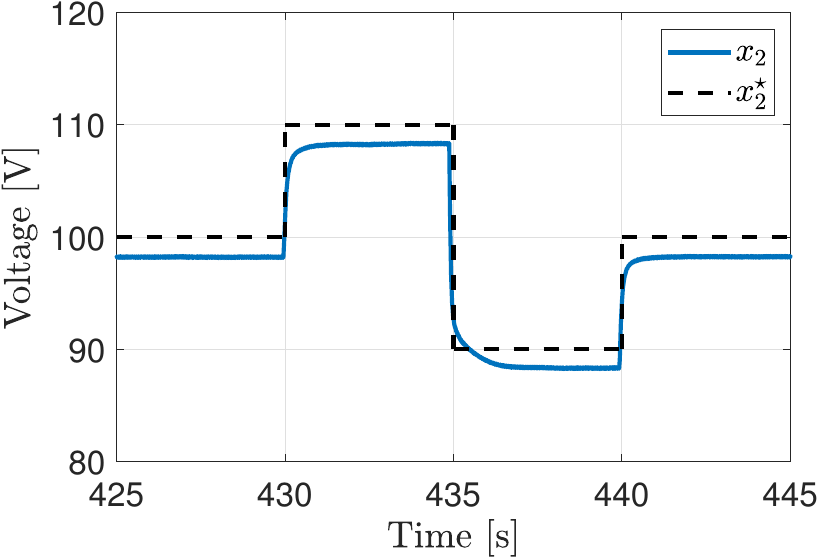}
	\caption{$x_2$ when $x_2^{\star}$ varies.}
	\label{x222}
\end{figure}

For subsystem $\Sigma_2$, the objective is to control the voltage $x_2$, as demonstrated earlier. The steady-state error is bounded by $\kappa_5$, $G$, and other system variables, as shown in the second property in proposition 1 . To test the controller designed for this subsystem, initially with zero load, reference changes were made in $x_2^{\star}$, considering the values of $\kappa_4$ and $\kappa_5$ from table \ref{tab:controller_2}. Fig. \ref{x222} shows the dynamic response of the signal $x_2$, from which it can be seen that the effect of $x_2^{\star}$ do not noticeably influence the error.

\begin{figure}
	\centering
	\includegraphics[width=0.5\textwidth]{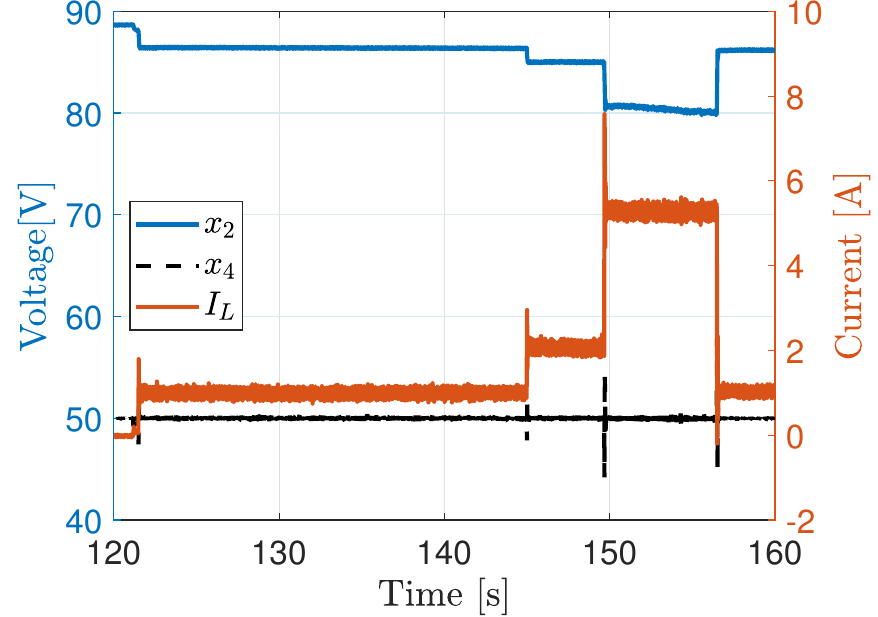}
	\caption{Current load effect in the controlled variable $x_2$ and $x_4$.}
	\label{il_exp}
\end{figure}

Another factor that affects the steady-state error is the system load, impacting variables $x_1$ and $x_3$. To observe this effect, current steps $I_L$ were applied, as depicted in Fig. \ref{il_exp}. Increasing $I_L$ (and consequently, $x_3$ and $x_1$) leads to an increase in error. It is crucial to be mindful of this, as the choice of $\kappa_5$ should consider the maximum current that the supercapacitor can deliver (or absorb). This ensures that the selected $\kappa_5$ value always maintains a voltage level higher than that of the supercapacitor, failure to achieve this can disrupt the system's operation and compromise the boosting and bucking capabilities of the converter.

\begin{figure}
	\centering
	\includegraphics[width=0.5\textwidth]{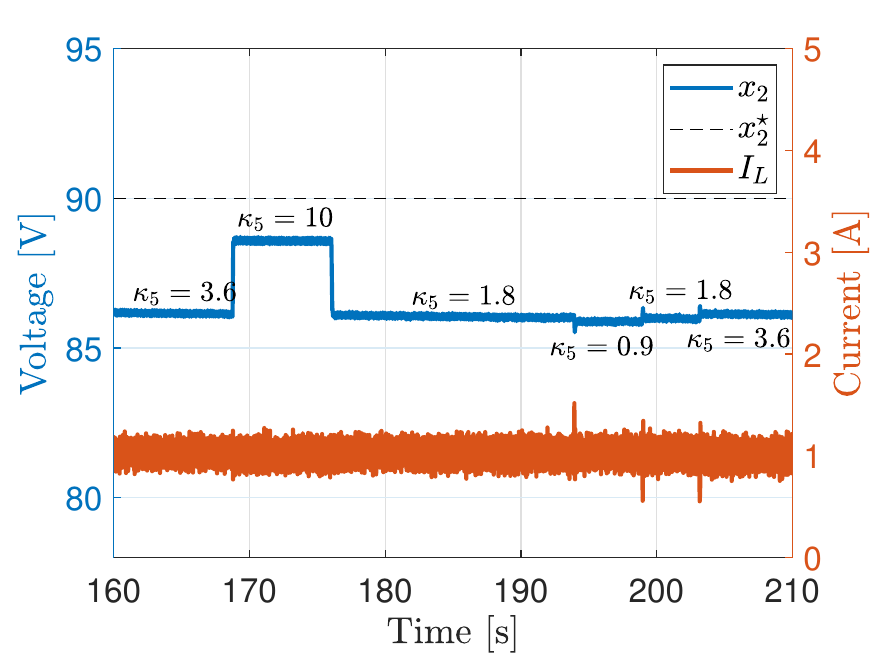}
	\caption{$\kappa_5$ change effect in the controlled variable $x_2$.}
	\label{k5_exp}
\end{figure}

We underscore the fact that that the purpose of this system is to regulate the tension $x_4$ and, consequently, the power supplied to the load. In this context, variations in $x_2$ do not affect the objective, as long as the tension $x_4$ successfully rejects disturbances and continues to regulate the tension to the desired value, as shown in Fig. \ref{il_exp}. Furthermore, it is important to observe the impact of $\kappa_5$ on the error between $x_2$ and $x_2^{\star}$, as illustrated in Fig. \ref{k5_exp}. Maintaining a constant value of $I_L$ at $\SI{1}{\ampere}$, we varied the value of $\kappa_5$. Initially, we increased $\kappa_5$ from 3.6 to 10, resulting in a decrease in error. Subsequently, we reverted to the  $\kappa_5=1.8$ value, then reduced it to 0.9, causing an increase in error. Lastly, we increased the value of $\kappa_5$, to 3.6, observing a  decrease in error.

\begin{figure}
	\centering
	\includegraphics[width=0.5\textwidth]{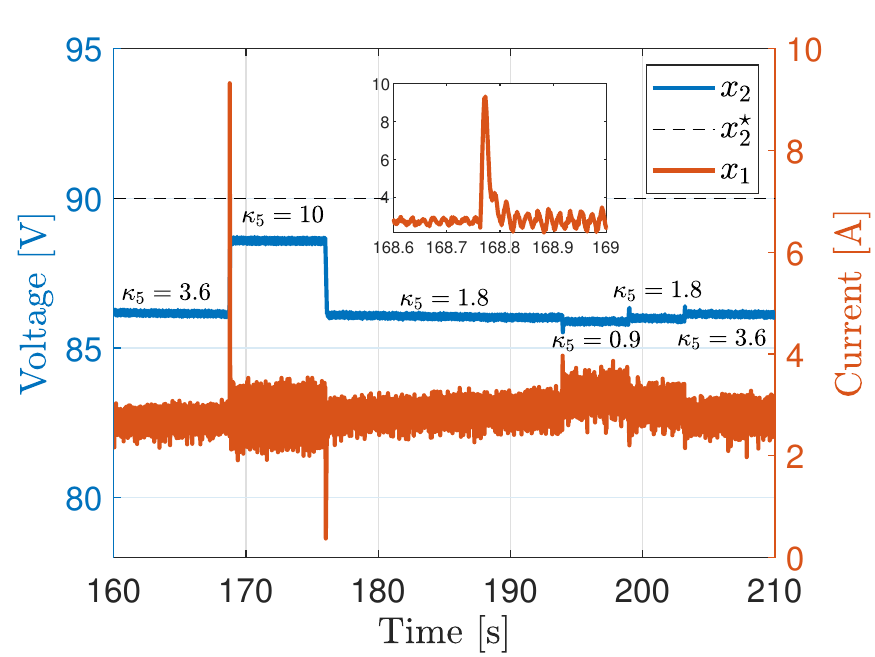}
	\caption{Impact of the value of $\kappa_5$ on $x_1$.}
	\label{k52_exp}
\end{figure}

Based on the aforementioned observation, it could be thought that the ideal choice would be to take $\kappa_5$ as large as possible when $I_L>0$. However, the selection of this parameter has an effect on the dynamics of $x_1$: the higher the value is, the greater the overshoot of the $x_1$ will be. Namely, there is a compromise between the regulation error of $x_2$ and the transient behaviour of $x_1$ when selecting $\kappa_1$. The impact of different $\kappa_5$ values on the current $x_1$ is shown in Fig. \ref{k52_exp}. 

\begin{figure}
    \centering
    \includegraphics[width=0.6\textwidth]{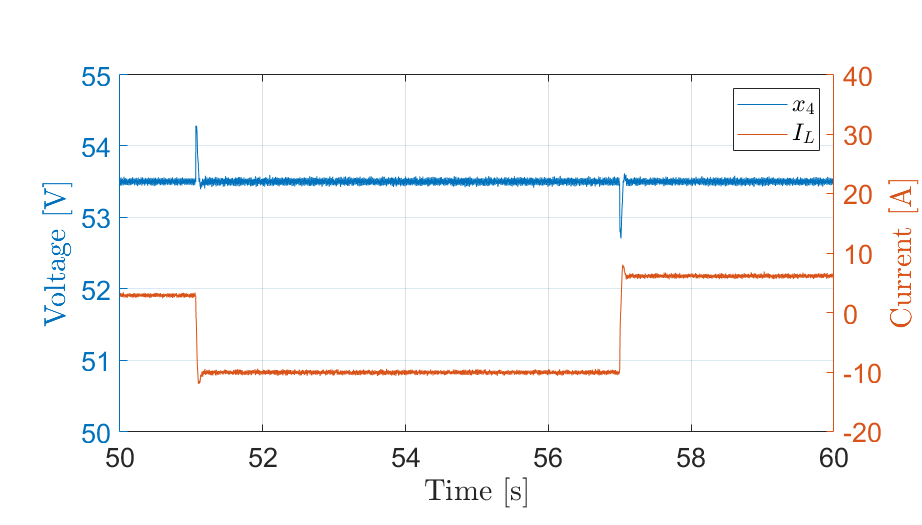}
    \caption{Effect on Load variation on $x_4$ regulated value.}
    \label{fig:il_minus}
\end{figure}

As a final test, we evaluate the performance of the voltage regulation in $x_4$ when $I_L$ is negative. As one can see from Fig. \ref{fig:il_minus},  regulation is achieved regardless the sign of $I_L$.


\section{Concluding Remarks and Future Research}
\lab{sec7}
%
In this work, a nonlinear control ensuring voltage regulation of a power system was proposed. The system is made up of a DC-DC converter which processes energy in a bidirectional way, a supercapacitor acting as an auxiliary storage, and a current source which supplies energy to---or draws it from---the energy storage. The design of the proposed controller exploits the system structure to derive a control law that regulates the voltage at one of the converter ports while the other signals remain bounded. Tuning its gains proves to be straightforward, as the dynamic behavior is guaranteed for a wide range of tuning parameters and the performance can be anticipated through simulations, prior to the actual implementation in a experimental setup. One of its gains, $\kappa_5$, holds significant importance, as it directly influences the error in the controlled voltage $x_2$. Moreover, as discussed in Subsection \ref{subsec43}, special care must be taken in the choice of this parameter in the energy flow to the supercapacitor changes sign. Emphasis is placed on the need to carry out a joint design between the power system and the control strategy, thereby bounding the limits of current supplied or absorbed by the converter and avoiding adverse effects such as damage to components due to overvoltages.

Regulate the value of $x_4$ accurately is one of the main goals of the controller, and this is achieved seamlessly. This demonstrates the control strategy's capability to be utilized in large range of DC-DC converters widely used in renewable energy systems or electric vehicles, where both, voltage step-up and step-down features, along with bidirectional current flow are required. This allows the use of energy storage sources whose voltage range operates both above and below the respective application, such as commercially available high-power supercapacitors or fuel cells.

Future work is oriented to the contrast of this regulation method with classical control techniques found in the literature, analyzing their respective advantages and disadvantages. Additionally, it is imperative to test this control scheme in multi-port systems, such as electric powertrains incorporating fuel cells and batteries, energy storage systems, and hydrogen production processes.

\bibliographystyle{plain}        
\bibliography{references}        
                            
\end{document}